\documentclass[cits]{PoS}

\newcommand{\beq}{\begin{eqnarray}}
\newcommand{\eeq}{\end{eqnarray}}

\newcommand{\D}{\mathcal{D}}

\newcommand{\os}{\overline{\s}}

\newcommand{\tr}{\mbox{Tr}}

\newcommand{\s}{\sigma}

\renewcommand{\D}{\Delta}

\newcommand{\oh}{\frac{1}{2}}

\newcommand{\non}{\nonumber}

\newcommand{\ra}{\rightarrow}

\newcommand{\rf}[1]{(\ref{#1})}

\title{k-string tensions and their large-N dependence  \thanks{CERN-PH-TH-2011-263} }

\ShortTitle{k-string tensions}

\author{\speaker{J. Greensite}\thanks{Supported in part by 
the U.S.\ Department of Energy under Grant No.\ DE-FG03-92ER40711.}\\
Physics and Astronomy Dept., San Francisco State
University, San Francisco, CA~94132, USA\\
        E-mail: \email{greensit@sfsu.edu}}

\author{B. Lucini \thanks{Supported by the Royal Society through the University 
Research Fellowship scheme and by STFC under contract ST/G000506/1.} \\
College of Science, Swansea University, Singleton Park, Swansea SA2 8PP, UK \\
E-mail: \email{B.Lucini@swansea.ac.uk}}

\author{A. Patella \thanks{Supported in part by the European Community - Research Infrastructure Action
under the FP7 ``Capacities'' Specific Programme, project ``HadronPhysics2.''}\\
CERN, Physics Department, 1211 Geneva 23, Switzerland \\
E-mail: \email{agostino.patella@cern.ch}}

\abstract{We consider whether the $1/N$ corrections to $k$-string tensions must begin at order $1/N^2$, as in the Sine Law, or whether odd powers of $1/N$, as in Casimir Scaling, are also acceptable. The issue is important because different models of confinement differ in their predictions for the representation-dependence of $k$-string tensions, and corrections involving odd powers of $1/N$ would seem to be ruled out by the large-$N$ expansion. We show, however, that $k$-string tensions may, in fact, have leading $1/N$ corrections, and consistency with the large-$N$ expansion, in the open string sector, is achieved by an exact pairwise cancellation among terms involving odd powers of $1/N$ in particular combinations of Wilson loops. It is shown how these cancellations come about in a concrete example, namely, strong coupling lattice gauge theory with the heat-kernel action, in which $k$-string tensions follow the Casimir scaling rule.}

\FullConference{ The XXIX International Symposium on Lattice Field Theory - Lattice 2011\\
July 10-16, 2011\\
Squaw Valley, Lake Tahoe, California}

\begin{document}

\section{Introduction}

The static potential of two heavy quarks in group representation $r$ and $N$-ality $k$ depends, asymptotically, only on the $N$-ality of the representation, i.e.\ $V(R)=\s_k R$, and these asymptotic string tensions are known as ``$k$-string tensions.''  Different confinement mechanisms lead to different predictions for $\s_k$,  the two most common of which are
\beq
\s_k = \s_F \times \left\{ \begin{array}{cl} 
         {k (N-k) \over (N-1)} & \mbox{Casimir Scaling} \cr \\
         {\sin\left({ \pi k \over N}\right) \over \sin\left({ \pi \over N}\right) }& \mbox{Sine Law} \end{array} \right. \; ,
\label{laws}
\eeq         
 where $\s_F$ is the string tension in the fundamental representation.  The Sine Law is found in certain supersymmetric models \cite{Douglas:1995nw}, in MQCD \cite{Hanany:1997hr}, in some AdS/CFT-inspired models \cite{Herzog:2001fq},
 and in certain versions of large-$N$ volume reduction where abelian dominance is assumed \cite{Armoni:2011dw}.  Casimir scaling, as originally proposed, means that 
 $\s_r = \s_F C_r/C_F$, where  $C_r$  is the quadratic Casimir in representation $r$.  This relation can be derived from the ``dimensional reduction'' form of the Yang-Mills vacuum wavefunctional \cite{Greensite:2007ij}, from the stochastic vacuum picture \cite{Di Giacomo:2000va},
from certain supersymmetric dual models \cite{Auzzi:2008ep}, and in the gauge-adjoint Higgs model in $D=3$ dimensions \cite{Antonov:2003tz}.  In these pictures, Casimir scaling should hold up to the distance where the quarks are screened by gluons.  Beyond that scale,  gluons screen the quark charge down to the representation of the same $N$-ality with the lowest dimensionality (smallest string tension), and then $C_r$ should be replaced by the Casimir of that lowest-dimension representation.  This gives us the Casimir scaling prediction shown in \rf{laws}.     
 
    Of course, neither behavior has to be exact.  There could be corrections.  But which prediction is closest to the truth, in ordinary, non-supersymmetric gauge theory?  That might tell us something about the nature of the confinement mechanism.  Armoni and Shifman, in ref.\ \cite{Armoni:2003nz} (see also Strassler \cite{Strassler:1997ny}), put forward a simple and powerful argument that Casimir scaling cannot be correct, because it conflicts with the large-$N$ expansion. Consider a product of $k$ rectangular $R \times T$ Wilson loops, with $T \gg R$, and let $U(R,T)$ be a Wilson loop holonomy around the rectangular contour.  On general grounds
\beq
            {1\over N^k} \langle (\tr[U(R,T)])^k \rangle = \sum_n a_n e^{-E_k^n(R) T} \; .
\label{sum_n}
\eeq
At large $R$, and as $T \ra \infty$ , the sum is dominated by the lowest energy $E_k^{min}(R) \sim \s_k R$, and therefore
\beq
{1\over N^k} \langle (\tr[U(R,T)])^k \rangle \ra a_{min} e^{-\s_k R T} \; .
\label{general}
\eeq   
On the other hand, according to the large-N expansion
\beq
{1\over N^k} \langle (\tr[U(R,T)])^k \rangle   = \left \langle {1\over N} \tr[U(R,T)]\right\rangle^k 
     + O\left({1\over N^2}\right) \; .
\label{largeN}
\eeq  
Inserting \rf{general} into \rf{largeN}, and taking the logarithm of both sides, we find from the $T\ra \infty$ limit
that  $\s_k = k \s_F  +$ powers of $1/N^2$.   On the other hand
\beq
\s_k = \s_F \times \left\{ \begin{array}{cl} 
         k - {k(k-1) \over N} + ... & \mbox{Casimir Scaling} \cr \\
         k - {\pi^2 (k^3-k) \over 6}{1\over N^2}+...& \mbox{Sine Law} \end{array} \right. \; .
\eeq     
In Casimir scaling, the leading correction starts at $1/N$ rather than $1/N^2$, and therefore Casimir scaling appears to be ruled out by large-$N$ expansion.

    However, lattice simulations do not seem to support this conclusion.   Very accurate simulations by Bringoltz and Teper in 2+1 dimensions \cite{Bringoltz:2008nd} strongly indicate $1/N$ , rather than $1/N^2$, leading corrections to the 
$k$-string tensions.  Thus we must ask if there is a loophole in the Armoni-Shifman argument.   A more detailed exposition of our analysis below is contained in ref.\ \cite{Greensite:2011gg},  and some of our conclusions were also anticipated in \cite{KorthalsAltes:2005ph}.
    
\section{The Cosh Argument, and a Strong-Coupling Example}

    We begin with a seemingly trivial question:  Does $\log\cosh(x)$ have an expansion in both even and odd powers of 
 $x$, or only even powers?  The answer, of course, is that the expansion is only in even powers, i.e.
\beq
    \cosh(x) &=& \sum_{n=0}^\infty {x^{2n} \over (2n)!} \\
\log \cosh(x) &=& \oh x^2 - {1\over 12} x^4 + {1\over 45} x^6 - {17 \over 2520} x^8  + ...        
\eeq    
However, suppose we consider  $|x| \gg 1$.  Then  $\cosh(x) \approx  {1\over 2} e^{|x|}$, and therefore 
$\log \cosh(x) \approx |x|$.  So if we drop one of the exponentials, which we can do if $|x| \gg 1$,  a power linear in $x$ turns up.
    
   To see the relevance to the Armoni-Shifman argument, return to eq.\ \rf{sum_n}, and suppose that the $E_k^n$ terms only differ by O($1/N$).   Then to retain only the  $\exp[-E_k^{min}(R) T]$ term, as in \rf{general}, it is necessary to keep $N$ fixed (but as large as desired), and then take the $T \ra \infty$ limit.   However, we get a different answer by keeping $T$ fixed (as large as desired) and then taking the $N \ra \infty$  limit.  In that case it is never admissible to truncate the sum in
\rf{sum_n} to a single term, and, as we will see, this has important consequences for the large-$N$ expansion.  The point here is that the $T \ra \infty$ and $N\ra \infty$ limits do not commute, and to get the usual large-$N$ expansion, \emph{the  large-N limit must come first}.  That is the limit corresponding to the small $x$ expansion of  $\log[\cosh(x)]$, to which both exponentials contribute.
   
   The lattice strong-coupling expansion of the heat-kernel action provides us with an explicit example of a theory with both Casimir scaling of the string tensions, and a standard large-$N$ expansion.  The heat-kernel action is derived by starting with the lattice Hamiltonian $H = g^2 \sum_{l} E^a_l E^a_l +  \sum_p V[U(p)]$, and choosing $V[U]$ by requiring that $\exp[-Ha]$ is the transfer matrix of some Euclidean theory on a hypercubic lattice.  The corresponding Euclidean theory turns out to be
\beq
   e^{-S} = \prod_p \sum_{R_p} d_{R_p} e^{-g^2 C_{R_p}/2} \chi_{R_p}[U(p)] \; ,
\eeq
where $d_R$ is the dimension of the representation and $C_R$ is the quadratic Casimir, with $\chi_R[U] = \tr_R[U]$ the $SU(N)$ group character. The product is over plaquettes $p$, and the sum is over group representations.  In this theory a planar Wilson loop, to leading order in the strong-coupling expansion, is given by  $\langle \chi_r[U(R,T)]\rangle = d_r \exp[-\s_r RT]$, where $\s_r$ satisfies the Casimir scaling formula
$\s_r = (C_r/C_F)\s_F$.  The lattice strong coupling expansion is known to be consistent with the large-$N$ expansion \cite{Yuri}, so how can this formula be correct, i.e.\ consistent with \rf{largeN}?

\begin{figure}
\begin{center}
\includegraphics[height=2.0cm]{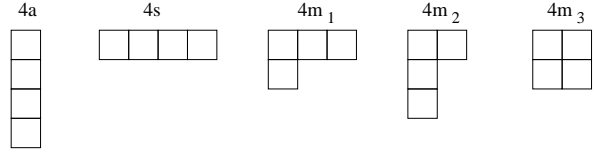}    
\end{center}
\caption{Young tableau for the $k=4$ calculation.  Representation pairs $4a,4s$, and $4m_1,4m_2$ are 
RC-conjugate (see below), while $4m_3$ is RC self-conjugate.}
\label{young3}
\end{figure}

   What happens is an (apparently) miraculous cancellation.  Let us consider $k=4$, for example, and let $U(C)$ be a Wilson loop holonomy around a planar loop $C$ bounding a minimal area $A$.  From standard group theory, expanding a product into irreducible representations,
\beq
(\tr ~ U(C))^4 =\tr_{4s}U(C) + \tr_{4a}U(C) + 3(\tr_{4m_1}U(C) + \tr_{4m_2}U(C)) + 2\tr_{4m_3}U(C)    \; ,
\eeq           
where the Young tableaux for the fully antisymmetric ($4a$), fully symmetric ($4s$), and mixed ($4m_{1-3}$) representations are displayed in Fig.\ \ref{young3}.  Introducing string tensions
\beq
\os &\equiv& \oh (\s_{4a} + \s_{4s})
 = \oh (\s_{4m_1} + \s_{4m_2})
 = \s_{4m_3} 
 = \left( 2N - {8\over N} \right) {2N \over N^2-1} \s \\
 &=& \left(4 + \mbox{even powers of }{1\over N}\right) \s  \; , \\
\D \s_{as} &\equiv& \s_{4s} - \s_{4a} = {24N \over N^2 -1} \s ~~ ,
~~
\D \s_{12} \equiv \s_{4m_1} - \s_{4m_2} = {8N \over N^2 -1} \s \; ,
\eeq
we find
\beq
&& {1\over N^4}\langle (\tr[U(C)])^4 \rangle = e^{-\os A} \left\{ {1\over 12}\left(1 + {11\over N^2}\right)\cosh(\oh\D \s_{as} A) 
 - {1\over 2}\left({1\over N} + {1\over N^3}\right)\sinh(\oh\D \s_{as} A) \right.
\non \\
  && \qquad + \left.  {3\over 4}\left(1-{1\over N^2}\right) \cosh(\oh\D \s_{12} A) 
- {3\over 2}\left({1\over N}-{1\over N^3}\right)\sinh(\oh\D \s_{12} A)
+ {1\over 6}\left(1-{1\over N^2}\right) \right\}  \; ,
\eeq
which is in perfect agreement with large-$N$ expectations, as there only even powers of $1/N$ on the right-hand side.  What has happened is that odd powers of $1/N$ have contrived to perfectly cancel, among the $4s, 4a, 4m_1, 4m_2, 4m_3$ representations.  A closer inspection shows that there is actually pairwise cancellation between the $4s,4a$ pair of representations, and the $4m_1,4m_2$ pair.

\begin{figure}
\begin{center}
\includegraphics[height=2.0cm]{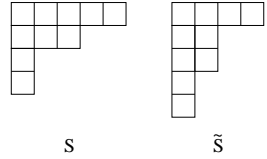}    
\end{center}
\caption{An example of an RC-conjugate pair of Young tableaux $(S,\widetilde{S})$.}
\label{RCconj}
\end{figure}

    This cancellation is not a coincidence, peculiar to $k=4$; it can be proven that to occur for any $k$.  We first introduce the notion of RC (row-column) conjugate representations, whose Young Tableaux are related by interchanging rows and columns (see Fig.\ \ref{RCconj}).  Then decompose the product representation into a sum of irreducible representations.  It turns out that the multiplicities of RC-conjugate representations are identical, so we may write
\beq
{1\over N^k}\langle \left( \tr U(C) \right)^k \rangle
= {1\over N^k} \sum_i g_i^k \left( \langle \tr_{R_i^k} U(C)
  \rangle + \langle \tr_{\widetilde{R}_i^k} U(C) \rangle\right)  \; ,
\label{irred}
\eeq
where $g_i^k$ is the multiplicity of representations $R_i^k$ and $\widetilde{R}_i^k$ or, in the case of an RC self-conjugate representation $R_i^k=\widetilde{R}_i^k$, it is half the multiplicity.
Now in the heat kernel action to leading order, we have already seen that {${1\over N^k} \langle \tr_r[U(R,T)]\rangle = {d_r\over N^k} \exp\left[-{C_r \over C_F} \s_F RT\right]$}.  Both $d_r$ and $C_r/C_F$ are functions of $N$, and it can proven that
\beq
\left[ {g_{\widetilde{R}} d_{\widetilde{R}} \over N^k} \right](N) &=&  
\left[ {g_R d_R \over N^k}\right](-N) \\
\left[{C_{\widetilde{R}} \over C_F}\right](N) &=& \left[{C_R \over C_F}\right](-N)     \; .
\eeq
As a result, ${1\over N^k}\langle \left( \tr U \right)^k \rangle
= \sum_i \left[ F_i\left({1\over N}\right) + F_i\left(-{1\over N}\right) \right]$. In other words, odd powers of $1/N$ cancel pairwise, among RC-conjugate representations.  This is how Casimir scaling is consistent with the $1/N$ expansion.

   We have not yet considered higher-order diagrams, such as a tube of plaquettes surrounding the loop, which are of essential importance in string-breaking/color-screening processes.  These can be shown, in various examples, not to alter the cancellation of odd powers of $1/N$.

\section{Beyond Strong-Coupling: A Theorem}

   It turns out that pairwise cancellations among RC-conjugate representations are general!  Examples are helpful, but the cancellation phenomenon we have discussed does not really depend on the heat-kernel action, Casimir scaling, or the strong-coupling expansion.

Using only group representation theory, and the standard large-$N$ expansion which tells us that 
$\langle \tr[U^p(C)]\rangle = N\times $ (a power series in $1/N^2$), 
we are able to prove that if $S$ and $\widetilde{S}$ are RC-conjugate representations 
with $k$ boxes in their corresponding Young Tableaux, then
\beq
\frac{\langle \tr_S ~ U(C) \rangle + \langle \tr_{\widetilde{S}} ~ U(C) \rangle }{N^k}
~~~\mbox{is an even function of}~~~{1\over N} \; ,
\eeq
and therefore has a power series expansion in powers of $1/N^2$.  The proof is non-trivial, and can be found in Appendix 
A of \cite{Greensite:2011gg}.    It follows that pairwise cancellation of odd powers of $1/N$ in \rf{irred} \emph{always} works, whether or not the string tensions exactly obey Casimir scaling.  It is this theorem which allows Casimir scaling, or some other rule for the string tensions which has a leading correction of order $1/N$, to be compatible with the large-$N$ expansion.

\section{Closed Strings}
     Let us now consider states created by Wilson lines winding through the periodic lattice in a spatial direction.  These closed-string states are sometimes called ``torelons.''  Unlike states created by timelike Wilson loop operators, the closed  string spectrum cannot be classified by the representation of the quarks at the ends, but only by the $N$-ality of the torelon creation operators. This brings in some new features, and again emphasizes the non-commutativity of large-$N$ and large distance limits.  
     
     The spectrum can be computed from the eigenvalues of the appropriate transfer matrix.  Let us consider the strong-coupling calculation in the $N$-ality $k=2$ sector, and compute the transfer matrix in the subspace of states spanned by torelon lines in the $2s$ and $2a$ (symmetric and antisymmetric) representations, which wind through the periodic lattice of extension $r$ in the $z$-direction.   Denote the lower and higher energy eigenvalues, obtained by diagonalizing the $2\times 2$ transfer matrix, as $E_L(r)$ and $E_H(r)$.  The explicit form of $E_L(r)$ and $E_H(r)$ can be found in 
\cite{Greensite:2011gg}, and they are complicated expressions in terms of torelon length $r$  and the $N$ of SU($N$),  but the important point is that, at fixed $r$ and taking the large-$N$ limit, these energies have a power series expansion in $1/N^2$.

      On the other hand, taking first the large $r$  limit at fixed $N$,  we get quite a different result, namely, $E_L(r) \approx 
 \s_{2a}r$, and $E_H(r)\approx \s_{2s}r$, where $\s_{2a,2s}$ are the Casimir-scaling string tensions of the $2a$ and $2s$ representations respectively, and these have leading order $1/N$ corrections.   Once again, it is matter of the order of limits.  If we take the large-$N$ limit first, then the energies have an expansion in terms of only even powers of $1/N$.   If, on the other hand, we take the large $r$ limit first, then leading corrections of order $1/N$ (or, strictly speaking, $1/|N|$)
 are possible.
 
     We may also ask if the torelon spectrum is degenerate in the large-$N$ limit.
A degeneracy is usually associated with a symmetry, which is not obviously present in this case. In fact we are able to show, again via strong-coupling examples (c.f.\ \cite{Greensite:2011gg}), that there is no degeneracy in the torelon spectrum in the  $N=\infty$ limit.
 
\section{Conclusions}
  
    We have shown that the large-$N$ expansion does not imply that $k$-string tensions necessarily have an expansion in only even powers of 1/$N$.   The fallacy in the argument to the contrary can be traced to the fact that the large-$N$ and large-distance limits do not commute.  Casimir scaling, in particular, can be compatible with the large-$N$ expansion. We also find that closed string sector has a non-degenerate spectrum, even at $N=\infty$ .

   The question of whether $k$-string tensions follow the Sine Law, or Casimir scaling, or something else is a dynamical issue.  It cannot be settled by large-$N$ counting alone.


\end{document}